\documentclass[12pt]{iopart}
\pdfoutput=1

\usepackage{graphicx}
\usepackage{dcolumn}
\usepackage{bm}
\usepackage{dsfont}
\usepackage{iopams}

\begin{document}

\title{Dissipative dynamics of a two - level system resonantly coupled to a
harmonic mode}
\author{Frederico Brito$^{1}$ and Amir O. Caldeira$^{2}$}

\address{$^{1}$IBM T. J. Watson Research Center, PO Box 218, Yorktown Heights, NY 10598 US,\\
$^{2}$Instituto de F{\'\i}sica {}``Gleb Wataghin''{}, Caixa Postal
6165, Universidade Estadual de Campinas-UNICAMP, 13083-970
Campinas, SP, Brazil}

\date{\today}

\begin{abstract}
We propose an approximation scheme to describe the dynamics of the
spin-boson model when the spectral density of the environment
shows a peak at a characteristic frequency $\Omega$ which can be very
close (or even equal) to the spin Zeeman frequency $\Delta$.
Mapping the problem onto a two-state system (TSS) coupled to a
harmonic oscillator (HO) with frequency $\omega_0$ we show that the
representation of displaced HO states provides an appropriate
basis to truncate the Hilbert space of the TSS-HO system and
therefore a better picture of the system dynamics. We derive an
effective Hamiltonian for the TSS-HO system, and show it furnishes
a very good approximation for the system dynamics even when its
two subsystems are moderately coupled. Finally, assuming the
regime of weak HO-bath coupling and low temperatures, we are able
to analytically evaluate the dissipative TSS dynamics.
\end{abstract}

\pacs{03.65.Yz, 03.67.Lx, 85.25.Dq}
\maketitle

\section{\label{introduction}Introduction}

As new experimental physics techniques allow us to reach minute
length scales, and milestones in control and precision of the
performed experiments are achieved, the study of open quantum
systems has become one of the most prominent areas of modern
physics. Moreover, both fundamental and practical aspects of the
systems under current investigation  help to foment a very broad
interest in the area. On the fundamental side, the possibility of
having a better comprehension of how classical physics emerges
from a quantum system or the realization of very peculiar quantum
mechanical states have attracted the interest of many researchers
towards the area. From the practical point of view, understanding
the main mechanisms leading to the loss of quantum coherence in
such systems constitutes a key feature, and one of the major
challenges for the physical implementation of, for example, a
quantum computer.

A very successful open quantum system which has been investigated
to date is a two-state system (TSS) coupled to a dissipative
environment. Despite its simplicity, the TSS dissipative dynamics
is  the paradigm of a wide variety of physical systems.
Superconducting devices containing Josephson junctions
\cite{Makhlin:2001p273}, few-electron semiconductor quantum dots
\cite{hanson:1217} and two-level atoms in optical cavities
\cite{cohen} are just few examples of  systems whose low-energy
level dynamics can be, in general, captured by a dissipative TSS.

The most studied model of a dissipative TSS is the well-known
spin-boson model \cite{Leggett:1987p5}, where the TSS is linearly
coupled to the coordinates of a bath of non-interacting
oscillators. The dissipative effects of the TSS phenomenological
environment are determined by the spectral density of the bath of
oscillators \cite{Caldeira:1983p1086}, which, in general, is
assumed to have a power-law behavior at low frequencies. Since it
seems that no general analytical solution can be obtained for such
a model, several approaches have been proposed to determine the
time evolution of the dissipative TSS . Those approaches basically
belong to two distinct approximation schemes: one considers a weak
TSS-bath coupling in the low-temperature regime, and the other
consists in performing an expansion in the tunnelling amplitude
$\Delta$ of the TSS states. The first approach is commonly
implemented by either using path-integral methods \cite{weiss} or
the Born-Markov approximation (also known as Bloch-Redfield
formalism)
\cite{Redfield,HartmannPhysRevE.61.R4687,divincenzo:035318}, where
only the lowest order terms in the TSS-bath coupling are taken
into account for the TSS dissipative dynamics. The other scheme
has been shown to be a fair approximation specially in the regimes
of strong damping and/or high temperatures, and also employs
path-integral methods within  the so-called \textit{noninteracting
blip approximation} (NIBA) as its main technique
\cite{Leggett:1987p5,weiss}.

An emerging problem in the area of dissipative TSSs is the one in
which  the bath effective spectral density $J_{eff}(\omega)$
presents a pronounced peak (resonance) at a characteristic
frequency $\Omega$. A typical example of such a case happens when
the energy scale of the device used to detect the state of the TSS
is comparable to that of its own regime of operation. Since the
device, or ``quantum detector'', also suffers from the dissipative
effects of the environment, the TSS-detector resonance constitutes
an efficient channel for decoherence processes to take place. For
example, as the energy scale of superconducting qubits reaches the
order of several GHz, this mechanism shows up on these systems
due to their coupling to the read-out dc-SQUIDs
\cite{Chiorescu:2003p285,Goorden:2004p307}. In addition to the
TSS-detector case, the presence of a structured bath with a sharp
peak has also been discussed in other pertinent contexts, {\it e.
g.}, electron  transfer in biological and chemical systems
 \cite{Garg:1985p303} and semiconductor quantum
dots \cite{HarryPhysRevB.70.195320}, to name just a few relevant
problems of this growing interest subject.

Indeed, it has been shown \cite{thorwart,wilhelm,Goorden:2004p307}
that the structure of the environment becomes an essential feature
of the dissipative TSS dynamics when its frequency is comparable
to the bath resonance frequency $\Omega$. Not only the decay rates
cannot be correctly determined by a perturbative approach in the
TSS-bath coupling, but such an approximation also fails to account
for the presence of additional resonances in the TSS dynamics. By
using an exact mapping \cite{Garg:1985p303} of the spin-boson
problem with a structured environment onto that of a TSS coupled
to a single HO system with frequency $\Omega$, which itself
interacts with a bath of oscillators having an Ohmic spectral
density, several techniques have been applied to the investigation
of the TSS dissipative dynamics. When the single HO is strongly
damped and/or coupled to a bath at high temperatures, the NIBA has
been employed to both TSS-HO weak \cite{Goorden:2004p307} and
strong \cite{NesiNJP} coupling cases.  On the other hand, the
numerically exact method of the quasi-adiabatic propagator
path-integral (QUAPI) \cite{thorwart,Goorden:2004p307}, as well as
a three-level system (3LS) Redfield theory,  have indicated the
failure of the perturbative approaches in the HO-bath weak
coupling and low temperatures regimes.

In this work, we present an approximation scheme to describe the
dynamics of the spin-boson problem with a structured bath having a
pronounced resonance at frequency $\Omega$. Following previous
works \cite{Garg:1985p303,thorwart,HarryPhysRevB.70.195320}, we
map this problem onto a TSS coupled to a single HO with frequency
$\omega_0$. In order to obtain an appropriate effective
Hamiltonian for the lowest-lying  energy states of the TSS-HO
system, we show that the representation of displaced HO states
whose displacement depends on the TSS state, provides a better
picture of the system dynamics, and consequently the appropriate
basis to truncate the Hilbert space of the TSS-HO system. We show
that the derived effective Hamiltonian furnishes a very good
approximation for a vast range of the physical parameters of the
system even when the TSS and the HO are moderately coupled  close
to the resonance. In addition, assuming the regime of weak HO-bath
coupling and low temperatures, we are able to analytically
evaluate the dissipative TSS spin dynamics.

This paper is organized as follows. In section \ref{model} we
present the model for the system, and derive the effective
Hamiltonian using the approximation scheme proposed above. By
varying the physical parameters of the problem, we compare its
predictions with: (1) an exact numerical calculation (for which we found to be
sufficient to consider a $N=18$ dimensional TSS-HO Hilbert space); and (2) the
simple truncation of the TSS-HO Hilbert space. Section
\ref{dynamics} contains the results obtained for the TSS-HO
dissipative dynamics assuming the regime of HO-bath weak-coupling
and low-temperatures. Finally, in section
 \ref{conclusion}, we present our concluding remarks.

\section{\label{model} The model}

Our starting point for describing the TSS dissipative dynamics in
the presence of a structured environment is the spin-boson
Hamiltonian \cite{Leggett:1987p5},
\begin{equation}
\hat{H}_{SB}=-\frac{\hbar}{2}\Delta\hat{\sigma}_x+\sum_k\hbar\omega_k
\hat{b}^\dagger_k\hat{b}_k+\hbar\hat{\sigma}_z\sum_k c_k(\hat{b}^\dagger_k
+\hat{b}_k),
\label{HSB}
\end{equation}
where $\hat{\sigma}_i$ are the Pauli matrices and $\hbar\Delta$
represents the tunnelling amplitude between the TSS states.

The bath of oscillators, introduced by the canonical bosonic
creation and annihilation operators $\hat{b}_k^\dagger$ and
$\hat{b}_k$, is assumed to have a spectral density
$J_{eff}(\omega)$ showing a Lorentzian peak at a characteristic
frequency $\Omega$.

It has been shown \cite{Garg:1985p303,HarryPhysRevB.70.195320}
that the proposed problem can be mapped onto a system comprised of
a TSS coupled to a single HO of frequency $\omega_0$, which itself
interacts with a bath of oscillators having a spectral density
$J(\omega)$ presenting a power-law frequency distribution. The
system Hamiltonian in this formulation is given by
\begin{eqnarray}
\lefteqn{\hat{H}=-\frac{\hbar}{2}\Delta\hat{\sigma}_x+\hbar\omega_0
\hat{a}^\dagger\hat{a}+\hbar\hat{\sigma}_z(g\hat{a}^\dagger+g^\ast\hat{a}) {}}\nonumber\\
&&\qquad\qquad+\sum_k\hbar\omega_k \hat{b}^\dagger_k\hat{b}_k+\hbar
(\hat{a}^\dagger+\hat{a})\sum_k \lambda_k(\hat{b}^\dagger_k+\hat{b}_k),
\label{TSSHO}
\end{eqnarray}
where $\hat{a}^\dagger$ and $\hat{a}$ are the creation and
annihilation operators of the single HO system, and $g$ stands for
the TSS-HO coupling constant. If $g$ is a real function, the
TSS-HO coupling has the form of a $``\hat{\sigma}_z$-coordinate''
interaction. On the other hand, if $g$ is a purely imaginary
function, the TSS-HO coupling becomes a
$``\hat{\sigma}_z$-momentum'' interaction.

From the system Hamiltonian (\ref{TSSHO}), one can observe that
the dissipative effects of the environment on the TSS dynamics
occur only indirectly through the TSS-HO interaction. Thus, in
this formulation, it is clear  that the TSS-HO coupling works as a
dissipative channel for the TSS. Such a channel can be enhanced or
suppressed depending on how strongly coupled the TSS and HO
systems are. This depends not only on the strength of the coupling
constant $g$, but also on how far from resonance these two
subsystems are. Indeed, if we assume a
$``\hat{\sigma}_z$-coordinate'' coupling, {\it i. e.}, $g=g^\ast$,
and an Ohmic spectral density $J(\omega)$,
\begin{equation}
J(\omega)=\frac{\pi}{2}\sum_k\lambda^2_k\delta(\omega-\omega_k)=
\kappa\omega e^{-\omega/\omega_c},
\label{ohmic}
\end{equation}
one can show \cite{Garg:1985p303} that the corresponding exact
mapping between the  Hamiltonians in (\ref{HSB}) and
(\ref{TSSHO})\footnote{Assuming a Lorentzian shape
$J_{eff}(\omega)=\frac{\pi}{2}\sum_k
c^2_k\delta(\omega-\omega_k)=\frac{2\alpha\omega\Omega^4}
{(\Omega^2-\omega^2)^2+(2\pi\kappa\omega\Omega)^2}$ for the
spectral density of the system Hamiltonian (\ref{HSB}).} occurs
when the single HO has its frequency given by the
$J_{eff}(\omega)$ resonance frequency $\Omega$, {\it i. e.},
$\omega_0=\Omega$. Consequently, the peak in the effective
spectral density of the bath ``seen'' by the TSS,
$J_{eff}(\omega)$, can be interpreted as a manifestation of a
definite resonance of one of the environment degrees of freedom.

Recently, Westfahl Jr. and collaborators
\cite{HarryPhysRevB.70.195320} have demonstrated that a
``$\hat{\sigma}_z$-momentum'' TSS-HO coupling, {\it i. e.},
$g=-g^\ast$, with a power-law frequency dependence for $J(\omega)$
($J(\omega)\propto \omega^s$), leads to a significant change of
the effective bath spectral density $J_{eff}(\omega)$ of the
system Hamiltonian (\ref{HSB}). Although $J_{eff}(\omega)$ still
shows a pronounced peak,   featuring an approximated Lorentzian
shape, its characteristic frequency $\Omega$ does not match the
single HO frequency $\omega_0$, {\it i. e.} $\Omega\neq\omega_0$.
Indeed, investigating the dissipative effects, due to the
spin-orbit mechanism, on the electronic spin dynamics trapped in
quantum dots, they have found that the characteristic resonance
frequency $\Omega$ predicted for the effective bath spectral
density $J_{eff}(\omega)$ is, in general, shifted to a value much
lower than that of the single HO, $\Omega\ll\omega_0$. In
addition, they have shown that such a resonance occurs in a
frequency regime which can be, in principle, of the order of the
Zeeman frequency of the spin, raising concerns about the
appropriate approach to correctly describe the dissipative spin
dynamics.

Thus, since the system Hamiltonian (\ref{TSSHO}) is capable of
describing the dissipative dynamics of a TSS coupled to a variety
of structured environments, from here onwards, we shall focus on
describing its features and properties. Unfortunately, despite its
simplicity, the Hamiltonian (\ref{TSSHO}) cannot be analytically
diagonalized.  However, since our final goal is to determine the
TSS dissipative dynamics in the regime of low temperatures,
$\hbar\Delta\gg k_B T$, we can concentrate on determining its
low-energy level spectrum. In addition, we are going to assume
that the HO dynamics is subject to weak dissipation, in such a way
that its states are weakly perturbed by the HO-bath coupling.

Under these considerations, in order to find a low-dimensional
effective Hamiltonian for the TSS-HO system, the natural procedure
is to perform a truncation of the HO Hilbert space, thus reducing
the TSS-HO Hilbert space dimensionality. The simplest method of
doing that consists in considering only the ground, $|0\rangle$,
and first excited, $|1\rangle$, states of the HO, leading to a
four-dimensional Hilbert space of the composite TSS-HO system.

\begin{figure}[t!]
\begin{center}\includegraphics[ width=0.95\columnwidth,
 keepaspectratio]{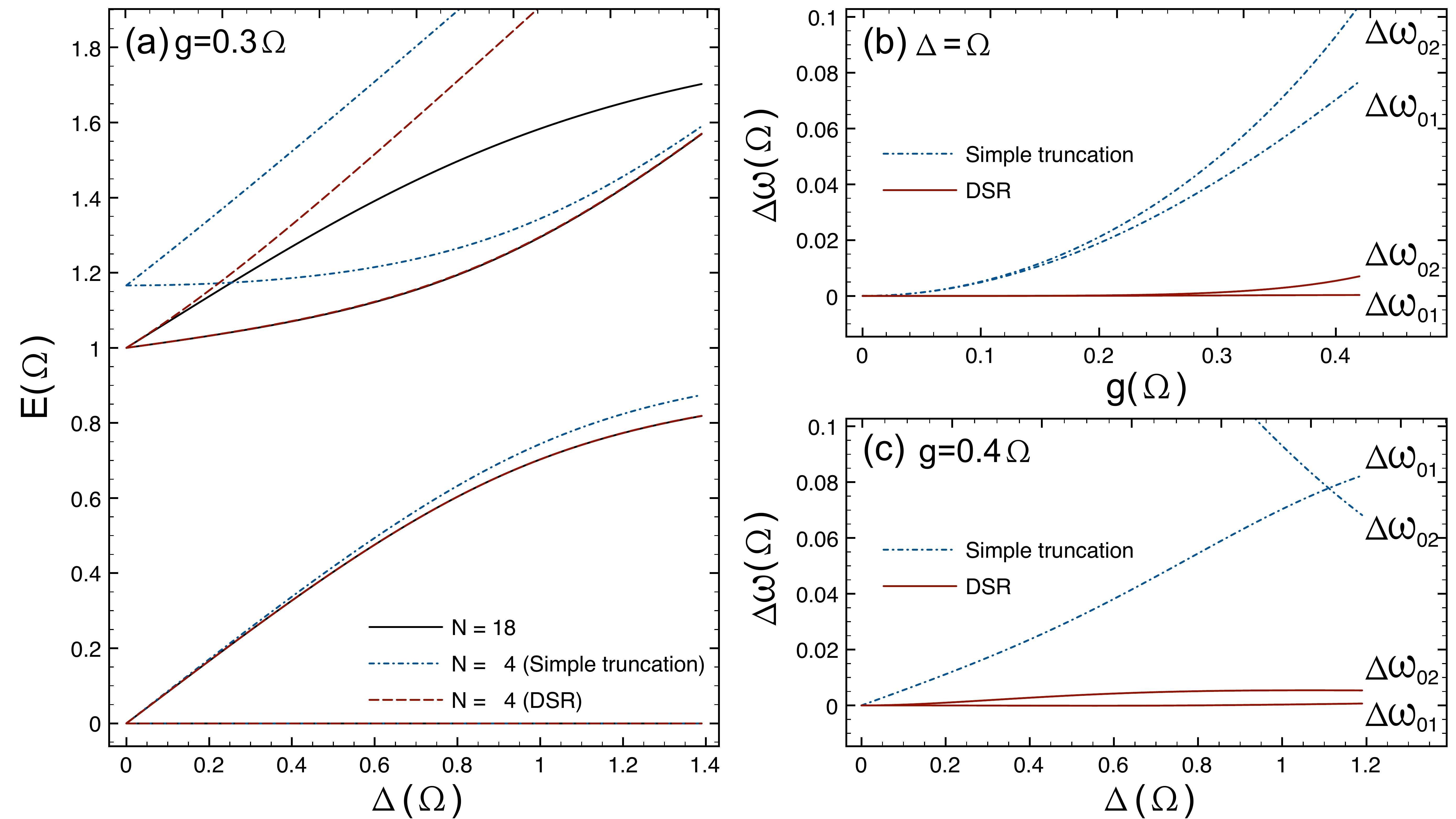}\end{center}
    \caption{(a) The four lowest levels of the TSS-HO system, as a
function of the tunnelling frequency $\Delta$, for the moderate
coupling case $g=0.3\Omega$, obtained through: numerical
calculation considering a $N=18$ dimensional Hilbert space (black
solid curve); analytical solution after performing a simple
truncation of the HO Hilbert space (blue dot-dashed curve); and
analytical solution of the effective Hamiltonian derived, Eq.
(\ref{Heff}), in the HO displaced state representation (DSR) (red
dashed curve). (The ground state is taken having $E=0$). (b) and
(c) present the $\omega_{01}$ and $\omega_{02}$  frequency
deviations,
$\Delta\omega_{0i}=\omega_{0i}^{ST,DSR}-\omega_{0i}^{(N=18)}$, for
the simple truncation approximation and the effective Hamiltonian
Eq. (\ref{Heff}), as a function of the TSS-HO coupling $g$ (for
the resonance case $\Delta=\Omega$), and the tunneling amplitude
$\Delta$ (assuming $g=0.4\Omega$).}
    \label{figure1}
\end{figure}

However, as depicted in Fig. \ref{figure1}, that approximation
fails as soon as the condition of a TSS-HO weak coupling,
$g\ll\Delta,\omega_0$ , is not satisfied.  Comparing with a
numerical calculation performed by taking into account a $N=18$
dimensional Hilbert space, we can see the system Bohr frequencies
$\omega_{01}$ and $\omega_{02}$ such that $\hbar\omega_{01}\equiv
E_1-E_0$ and $\hbar\omega_{02}\equiv E_2-E_0$, where $E_0,E_1$ and
$E_2$ respectively correspond to the eingenenergy of the ground,
first and second excited states, can deviate {\it circa} $10\%$
from the correct values, even for moderate TSS-HO couplings. Since
we expect a sharp resonance for the effective spectral density
$J_{eff}(\omega)$, those deviations can easily lead to much larger
errors when computing the TSS relaxation rates. Indeed, Thorwart
and collaborators \cite{thorwart} report deviations for the
relaxation rates of the order of $\approx20\%$ for the weak
coupling case $g=0.07$, when performing the simple truncation of
the HO Hilbert space.

The failure of the simple truncation of the HO Hilbert space for
cases other than the TSS-HO weak coupling, $g\ll\Delta,\Omega$ ,
was already expected, since it does not correctly take into
account the weight of the higher energy HO states in the spectral
decomposition of the lowest energy eigenstates of the Hamiltoninan
(\ref{TSSHO}). As one can observe from the latter equation, the
TSS-HO interaction term has the net effect of inducing
displacements of the HO in its phase space $\{q,p\}$, hence
leading to the mixing of the HO eigenstates of Eq.(\ref{TSSHO}).
So, the stronger the TSS-HO coupling is, the more important the
HO higher energy states become.

Therefore, in order to obtain a better picture of the lowest
energy states of Eq. (\ref{TSSHO}), we have to perform the
truncation of the HO Hilbert space in a representation where such
an effect is taken into account. In other words, it is the
displaced harmonic oscillator states which ought to be the
appropriate basis where one should perform the reduction of
Hilbert space of the TSS-HO system.

The change of representation can be done by introducing the
unitary HO displacement operator $\hat{D}(\alpha)=e^{\alpha
\hat{a}^\dagger-\alpha^\ast \hat{a}}$, which is very well-known in
the context of coherent states \cite{cohen}. There, among several
other implications, the parameter $\alpha$ is a $c$-number, such
that the transformation
$\hat{D}^\dagger(\alpha)\hat{a}\hat{D}(\alpha)=\hat{a}+\alpha$ has
the action of shifting the HO canonical position and momentum
operators by amounts proportional to,  ${\rm Re}(\alpha)$ and
${\rm Im}(\alpha)$, respectively. However, an important feature of
the Hamiltonian (\ref{TSSHO}) is the fact that the displacements
of the HO states  depend on the state of the TSS. Therefore, we
are led to introduce the unitary transformation
\begin{equation}
\hat{D}(s,\hat{\sigma}_z)\equiv
e^{(s\hat{a}^\dagger-s^\ast\hat{a})\hat{\sigma}_z},
\label{transformation}
\end{equation}
as the conditional displacement operator of Eq.(\ref{TSSHO}).
Here, we consider the displacement parameter $s$ as an {\it ad
hoc} parameter of our model, to be properly chosen in order to
obtain the best effective Hamiltonian possible.

Transformation (\ref{transformation}) is very similar to what is
known in the literature as the polaron transformation
\cite{Leggett:1987p5,weiss}. Indeed, if we had followed, by
analogy, the choice made for the polaron transformation, we would
have obtained $s=-g/\omega_0$. Although it correctly diagonalizes
the Hamiltonian (\ref{TSSHO}) when $\Delta=0$, this choice is
expected  to fail, as discussed in
 \cite{Brito:2008p1076}, when the tunnelling amplitude becomes
appreciable, since from its analytical solution one obtains
$\omega_{01}\rightarrow\infty$, instead of
$\omega_{01}\rightarrow\omega_0$, when $\Delta\rightarrow\infty$.

Thus, in order to make the best choice for the {\it ad hoc}
parameter $s$, we have to carefully observe which one  would
reflect the kind of dynamics we should expect for our TSS-HO
system. Performing the change of representation,
$\tilde{H}\equiv\hat{D}^\dagger\hat{H}\hat{D}$, and truncating the
HO Hilbert space, one finds
\begin{eqnarray}
\tilde{H}_{TSS-HO}=
-\frac{\hbar}{2}\tilde{\Delta}\hat{\sigma}_x\left(1-4|s|^2\hat{c}^\dagger\hat{c}\right)
+\hbar\omega_0\hat{c}^\dagger\hat{c}\nonumber\\
\qquad+\hbar\hat{c}^\dagger\left\{\hat{\sigma}_z(g+\omega_0s)+i\hat{\sigma}_y
s\tilde{\Delta}\right\}
+\hbar\hat{c}\left\{\hat{\sigma}_z(g^\ast+\omega_0s^\ast)-i\hat{\sigma}_y
s^\ast\tilde{\Delta}\right\}, \label{genDSR}
\end{eqnarray}
where, in order to stress the fact that we are now working with a
reduced Hilbert space, we have defined the operators
$\hat{c}\equiv|\tilde{0}\rangle\langle\tilde{1}|$ and
$\hat{c}^\dagger\equiv|\tilde{1}\rangle\langle\tilde{0}|$, in
which $|\tilde{0}\rangle$ and $|\tilde{1}\rangle$ represent,
respectively, the ground and first excited states of the displaced
HO. $\tilde{\Delta}$ represents the renormalized tunnelling
amplitude due to the TSS-HO coupling, which is given by
\begin{equation}
\tilde{\Delta}\equiv\Delta e^{-2|s|^2}. \label{delta}
\end{equation}

Without dissipation, the TSS-HO system should evolve in time in
such a way there would be  a persistent  exchange of a quantum of
energy between the two subsystems. Such a dynamical behaviour  has
a precise analogue in the problem of absorption (emission) of a
single photon by (from) an atom placed in an optical cavity.
There, the fundamental tool for the study of the system dynamics
is the so-called Jaynes-Cummings (JC) Hamiltonian. Inspecting Eq.
({\ref{genDSR}), one can see that the choice
\begin{equation}
s\equiv-\frac{g}{\omega_0+|\tilde{\Delta}|}, \label{parameters}
\end{equation}
leads us to the same form as a JC Hamiltonian. It worth noting
that one could use a variational calculation to determine the
value of the parameter $s$ for the polaron transformation Eq.
(\ref{transformation}). Indeed, by minimizing the free energy of
the system using the Bogoliubov-Peierls bound, Silbey and Harris
\cite{silbey:2615} could determine a transcendental equation for
$s$. For the unbiased case and $T=0$, that leads to a very similar
renormalized tunnelling amplitude to that found in Eq.
(\ref{delta}) (cf. Eq. (103) in the review \cite{goychuk})

Assuming
$\Delta\leq0$\footnote{For the case $\Delta\geq0$, the last term
of Eq. (\ref{Heff}) becomes
$(g\hat{c}^\dagger\hat{\sigma}^{(x)}_++g^\ast\hat{c}~\hat{\sigma}^{(x)}_-)$,
reflecting the fact that the state $|+\rangle$ has the lowest
Zeeman energy.}, the effective Hamiltonian in the HO displaced
state representation (DSR) obtained using Eq.(\ref{parameters}) in
Eq.(\ref{genDSR}) reads
\begin{eqnarray}
\tilde{H}_{DSR}\equiv-\frac{\hbar}{2}\tilde{\Delta}
\hat{\sigma}_x\left(1-\frac{4|g|^2}{(\omega_0+|\tilde{\Delta}|)^2}
\hat{c}^\dagger\hat{c}\right)+\hbar\omega_0\hat{c}^\dagger\hat{c}\nonumber\\
\qquad+\frac{2\hbar| \tilde{\Delta}|}{\omega_0+|\tilde{\Delta}|}(
g\hat{c}^\dagger\hat{\sigma}^{(x)}_-+g^\ast\hat{c}~\hat{\sigma}^{(x)}_+),
\label{Heff}
\end{eqnarray}
where we have introduced the ladder operators for the spin
$x-$component,
$\hat{\sigma}_\pm^{(x)}=\frac{1}{2}(\hat{\sigma}_z\mp
i\hat{\sigma}_y)$\footnote{The ladder $\hat{\sigma}_\pm^{(x)}$
operators obey the usual commutation relations:
$[\hat{\sigma}_x,\hat{\sigma}_\pm^{(x)}]= \pm
2\hat{\sigma}_{\pm}^{(x)}$ and
$[\hat{\sigma}_+^{(x)},\hat{\sigma}_-^{(x)}]=\hat{\sigma}_x$.
Their action on the eigenstates of $\hat{\sigma}_x$ are given by:
$\hat{\sigma}_+^{(x)}|-\rangle= |+\rangle$ and
$\hat{\sigma}_-^{(x)}|+\rangle=|-\rangle$, with
$(\hat{\sigma}_+^{(x)})^2=(\hat{\sigma}_-^{(x)})^2=0$.}. The
Hamiltonian (\ref{Heff}) can be diagonalized analytically,
yielding  eigenenergies given by
\begin{eqnarray}
E_0\equiv-\frac{\hbar}{2}\tilde{|\Delta|}, \qquad E_3
\equiv\frac{\hbar}{2}|\tilde{\Delta}|\left\{1-\frac{4|g|^2}
{(\omega_0+|\tilde{\Delta}|)^2}\right\}+\hbar\omega_0,~{\rm and}\label{E0E3}\\
E_{1,2}\equiv\frac{\hbar}{2}\left\{2\omega_0-
\frac{E_3+E_0}{\hbar}\mp\sqrt{\left(\frac{E_3-E_0}
{\hbar}-2\omega_0\right)^2+\frac{16|g|^2\tilde{\Delta}^2}
{(\omega_0+|\tilde{\Delta}|)^2}}\right\}.\label{E1E2}
\end{eqnarray}

The analysis of the obtained eigenenergies $E_i$, Eqs.
(\ref{E0E3}) and (\ref{E1E2}), reveals that the Hamiltonian
(\ref{Heff}) provides the correct results for certain limits of
the  parameters of Eq. (\ref{TSSHO}), namely: (1) as
$\omega_0\rightarrow\infty$, we obtain
$\omega_{01}\rightarrow|\Delta|$; (2) for
$|\Delta|\rightarrow\infty$, we have
$\omega_{01}\rightarrow\omega_0$; and for the resonant case
$\Delta=\omega_0$, with $g\ll\Delta,\omega_0$, we find
$\omega_{01}\approx\omega_0-g$ and $\omega_{02}\approx\omega_0+g$.
Moreover, as one can see from Fig. \ref{figure1}, the resulting
Hamiltonian (\ref{Heff})  gives a very good approximation of our
problem for a vast range of the physical parameters $\Delta$ and
$g$. In fact, the largest deviations we have found considering the
parameters illustrated in Fig. \ref{figure1} were
$\Delta\omega_{01}\approx 0.07\%$ and
$\Delta\omega_{02}\approx0.6\%$.  It is worth noting that, when
performing the simple truncation, this level of agreement can only
be met for the ground and first excited states if we consider a
$N=8$ dimensional Hilbert space. Indeed, for a $N=6$ dimensional
Hilbert space we found:
$\Delta\omega_{01}^{(N=6)}\approx20\Delta\omega_{01}^{DSR}$ and
$\Delta\omega_{02}^{(N=6)}\approx3\Delta\omega_{02}^{DSR}$, while
for the case $N=8$ we obtained:
$\Delta\omega_{01}^{(N=8)}\approx1.3\Delta\omega_{01}^{DSR}$ and
$\Delta\omega_{02}^{(N=8)}\approx0.34\Delta\omega_{02}^{DSR}$.

Finally, defining the states $|\pm\rangle$ as the eigenstates of
$\hat{\sigma}_x$, $\hat{\sigma}_x|\pm\rangle=\pm|\pm\rangle$, the
eigenstates of the Hamiltonian  (\ref{Heff}) can be written in the
form
\begin{eqnarray}
|e_0\rangle\equiv|-,\tilde{0}\rangle,\qquad |e_3\rangle\equiv|+
,\tilde{1}\rangle,\quad|e_{1}\rangle\equiv A|+,\tilde{0}\rangle
+B|-,\tilde{1}\rangle, ~{\rm and}\label{vecs1}\\
|e_{2}\rangle\equiv -B|+,\tilde{0}\rangle
+A|-,\tilde{1}\rangle,\label{vecs2}
\end{eqnarray}
where, $A\equiv
(E_0+E_2)/\sqrt{(E_0+E_2)^2+4|g|^2\tilde{\Delta}^2\hbar^2/(\omega_0+|\tilde{\Delta}|)^2}$
and $B^2=1-A^2$.

The eigenstates $|e_{1,2}\rangle$, Eqs. (\ref{vecs1}) and
(\ref{vecs2}), reveal the hybridization process occurring between
the TSS and HO states due to their coupling. Indeed, if $g=0$,
{\it i. e.}, the case of a decoupled TSS-HO system, the
eigenstates of Eq. (\ref{Heff}) are simple tensor product of the
TSS and HO eigenstates ($A=1$). Nevertheless, as one increases the
coupling $g$, the system eigenstates become more hybridized. At
the resonance condition, $\omega_0=\Delta$,  the eigenstates
$|e_{1,2}\rangle$ become nearly the maximally entangled TSS-HO
states, with $A\approx\frac{1}{\sqrt{2}}(1+O(g))$.

\section{Dissipative dynamics\label{dynamics}}

The evaluation of the dissipative TSS dynamics can be described,
for a weak HO-bath coupling and low temperatures, using the
Redfield equations \cite{weiss}
\begin{equation}
\dot{\tilde{\rho}}_{nm}(t)=-i\omega_{nm}\tilde{\rho}_{nm}(t)-
\sum_{k,l}R_{nmkl}\tilde{\rho}_{kl}(t),
\label{redfield}
\end{equation}
where the matrix elements $\tilde{\rho}_{nm}\equiv\langle n
|\tilde{\rho}|m\rangle$ are taken using the eigenstates of Eq.
(\ref{Heff}), and the Redfield tensors are given by
$R_{nmkl}=\delta_{lm}\sum_r\Gamma_{nrrk}^{(+)}+\delta_{nk}
\sum_r\Gamma_{lrrm}^{(-)}-\Gamma_{lmnk}^{(+)}-\Gamma_{lmnk}^{(-)}$,
with
\begin{eqnarray}
\Gamma_{lmnk}^{(+)}=\hbar^{-2}\int_0^\infty dt
e^{-i\omega_{nk}t}\langle\tilde{\cal{H}}_{I,lm}(t)\tilde{\cal{H}}_{I,nk}(0)\rangle,\\
\Gamma_{lmnk}^{(-)}=\hbar^{-2}\int_0^\infty dt
e^{-i\omega_{lm}t}\langle\tilde{\cal{H}}_{I,lm}(0)\tilde{\cal{H}}_{I,nk}(t)\rangle,
\end{eqnarray}
where $\tilde{\cal{H}}_{I}(t)=e^{i\tilde{H}_Rt/\hbar}\tilde{H}_I
e^{-i\tilde{H}_Rt/\hbar}$ is the interaction Hamiltonian in the
interaction picture, and the brackets represent thermal averages
over the bath degrees of freedom. Here, we denote $\tilde{H}_R$
and $\tilde{H}_I$ as the bath and HO-bath interaction Hamiltonians
obtained  from Eq. (\ref{TSSHO}) under the transformation
presented in the last section. The full Hamiltonian now reads
\begin{equation}
\tilde{H}=\tilde{H}_{DSR}+\sum_k\hbar\omega_k\hat{b}_k^\dagger\hat{b}_k+
\left(\hat{c}^\dagger+\hat{c}-\frac{2{\rm Re}(g)}{\omega_0+
|\tilde{\Delta}|}\hat{\sigma}_z\right)\sum_k\hbar\lambda_k(\hat{b}^\dagger+\hat{b}_k).
\label{fullDSR}
\end{equation}
where the second and third terms on the RHS of this equation are,
respectively, $\tilde{H}_R$ and $\tilde{H}_I$.  It is worth noting
that the term proportional to Re$(g)$ in $\tilde{H}_I$, Eq.
(\ref{fullDSR}), represents a correction naturally  introduced by
the scheme proposed, and leads to an effective interaction between
the TSS and bath mediated by the TSS-HO coupling. Had we followed
the simple truncation approximation, such a term would not be
present in the system dynamics. We have seen that this term can
represent a non-negligible correction to the system-bath
interaction, since it can reach fractions of few percent of the
latter, {\it e. g.}, for $g=0.1, \Delta=\omega_0=1$, we find
$||2{\rm
Re}(g)/(\omega_0+|\tilde{\Delta}|)||/||\hat{c}^\dagger+\hat{c}||\approx0.1$.

In order to be able to analyze the resonance case,
$\omega_0=\Delta$, as pointed out in Ref. \cite{thorwart}, we have
to go beyond the two-level approximation when using the Redfield
equations (\ref{redfield}). Indeed, since a two-level
approximation is not capable of taking into account the
hybridization of TSS-HO states, it fails to present the additional
resonances of the TSS dynamics. Thus, considering a three-level
system, one can determine the expectation value of $\sigma_z$,
$\sigma_z={\rm Tr}(\hat{\rho}(t)\hat{\sigma}_z)$, as
\begin{equation}
\sigma_z(t)\approx 2{\rm Re}\left(A\langle e_0|\tilde{\rho}|e_1
\rangle-B\langle e_0|\tilde{\rho}|e_2\rangle\right).\label{sigmaz}
\end{equation}
Performing the secular approximation \cite{Redfield}, we can
neglect the terms $R_{nmkl}$ of $\tilde{\rho}_{nm}$ for which
$\omega_{nm}\neq \omega_{kl}$. This leads us to the following set of coupled
differential equations
\begin{eqnarray}
\dot{\tilde{\rho}}_{01}\equiv\langle e_0|\dot{\tilde{\rho}}
|e_1\rangle\approx i\omega_{01}\tilde{\rho}_{01}-
\gamma_1\tilde{\rho}_{01}-\gamma_2\tilde{\rho}_{02},\label{rho01}\\
\dot{\tilde{\rho}}_{02}\equiv\langle e_0|\dot{\tilde{\rho}}
|e_2\rangle\approx i\omega_{02}\tilde{\rho}_{02}-
\gamma_3\tilde{\rho}_{01}-\gamma_4\tilde{\rho}_{02},\label{rho02}
\end{eqnarray}
where $\gamma_1= R_{0101}, \gamma_2=R_{0102},
\gamma_3=R_{0201},~{\rm and}~\gamma_4=R_{0202}$. Assuming that
$J(\omega)$ is a regular function in the complex plane, we can
analytically evaluate the Redfield rates, obtaining
\begin{eqnarray}
{\rm Re}~\Gamma_{lmnk}^{(+)}=(\tilde{h}_{lm}\tilde{h}_{mk})
J(|\omega_{nk}|)\frac{e^{-\beta\hbar\omega_{nk}/2}}
{\sinh(\beta\hbar|\omega_{nk}|/2)}\\
{\rm
Im}~\Gamma_{lmnk}^{(+)}=-\frac{2}{\pi}(\tilde{h}_{lm}\tilde{h}_{mk})
{\rm P}\int_0^\infty d\omega\frac{J(\omega)}
{\omega^2-\omega_{nk}^2}\left(\omega-\omega_{nk}\coth(\frac{\beta\hbar\omega}{2})\right)
\end{eqnarray}
where $\tilde{h}_{lm}\equiv\langle
l|(\hat{c}^\dagger+\hat{c}-\frac{2{\rm Re}(g)}
{\omega_0+|\tilde{\Delta}|}\hat{\sigma}_z)|m\rangle$. In addition,
using the relation
$\Gamma_{knml}^{(-)}=(\Gamma_{lmnk}^{(+)})^\ast$, we can determine
the Redfield tensors $R_{nmkl}$ in terms of the rates
$\Gamma_{lmnk}^{(+)}$.

From Eqs. (\ref{sigmaz}-\ref{rho02}), we find the spectral
decomposition of $\sigma_z$ as
\begin{equation}
\omega_{\pm}=\frac{\omega_{01}+\omega_{02}}{2}+
i\left(\frac{\gamma_1+\gamma_4}{2}\right)\pm
\sqrt{\left(\frac{\omega_{01}-\omega_{02}}{2}+
\frac{i}{2}(\gamma_1-\gamma_4)\right)^2-\gamma_2\gamma_3}
\label{damping}
\end{equation}
where Re$(\omega_\pm)$ are the main oscillation frequencies of the
$\sigma_z$ dynamics, and Im$(\omega_\pm)$ gives their respective
damping rates.

\begin{figure}[t!]
\begin{center}\includegraphics[ width=0.7\columnwidth,
 keepaspectratio]{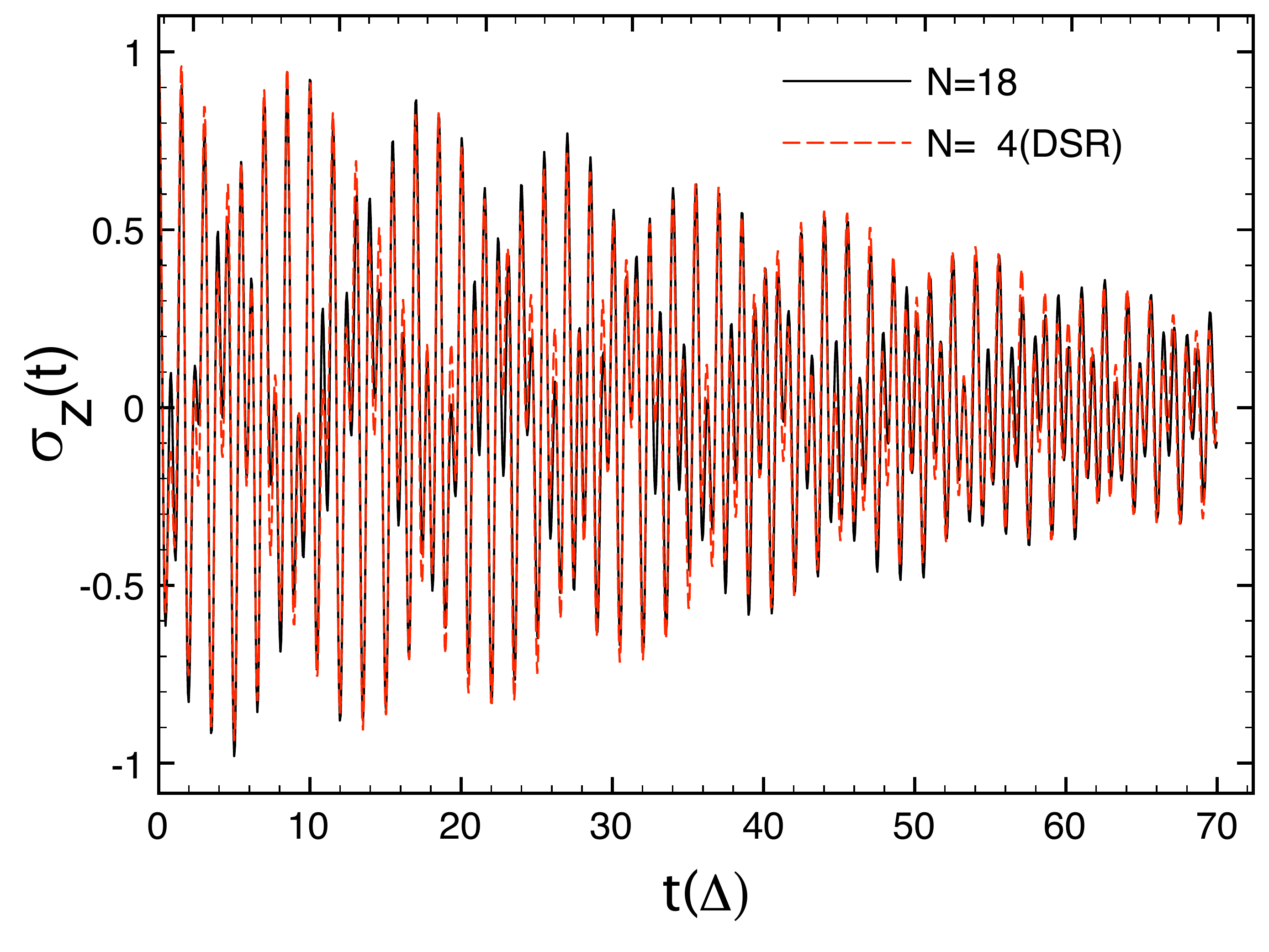}\end{center}
    \caption{$\sigma_z(t)$ dissipative dynamics for the resonant
case, $\Delta=\omega_0$, computed for an Ohmic spectral density,
Eq. (\ref{ohmic}), with $\kappa=0.02$, $\omega_c=10\Delta$ and
temperature $k_B T=0.1\hbar\Delta$. We assumed a
$\sigma_z-$coordinate TSS-HO coupling, with constant coupling
given by $g=0.3\Delta$. Two main oscillation frequencies
$\omega_{\pm}$, Eq. (\ref{damping}), are observed in the dynamics,
reflecting the hybridization of the TSS-HO states. Solid and
dotted lines represent, respectively, the $N=18$ dimensional
Hilbert space and our $N=4$ truncated space in the DSR.}
    \label{figure2}
\end{figure}

Figure \ref{figure2} presents a typical time evolution for
$\sigma_z$ for the resonant case using Eq. (\ref{damping}). Once
again, by comparing it to the dynamics presented by a larger
Hilbert space, $N=18$, we can verify a very good agreement of the
results obtained using the scheme developed in section \ref{model}
. As one can observe, the TSS dynamics presents a rich structure,
with the presence of two main oscillation frequencies, which are
due to the TSS-HO resonance. These results corroborated those
obtained in \cite{thorwart,wilhelm,Goorden:2004p307}.

\section{Conclusions\label{conclusion}}

In conclusion, we presented and discussed  a new approximation
scheme to describe the dynamics of the spin-boson model with a
structured environment. We studied the proposed problem by
exploring its exact mapping onto that of a TSS coupled to a single
HO of frequency $\omega_0$, which itself interacts with a bath of
oscillators.  We showed that, in order to find a low-dimensional
effective Hamiltonian for the TSS-HO system, the representation of
displaced HO states, DSR,  provides an appropriate basis to
truncate its Hilbert space. For that, we defined a conditional
displacement operator, where the shifting of the HO system was
dependent of the state of the TSS. In so doing we needed to
introduce an {\it ad hoc}  displacement parameter $s$  which
should be chosen in such a way  that the best effective
Hamiltonian could be obtained. Invoking the physics of an
atom-cavity system, we chose the parameter $s$ such that the
effective Hamiltonian had a Jaynes-Cummings form. By comparing our
numerical results with those considering a larger Hilbert space,
we showed that, in fact, the derived effective Hamiltonian gives
an excellent approximation of the original problem for a vast
range of its physical parameters even when the TSS and the HO are
moderately coupled close to the resonance. Actually, we showed
that in order to achieve the same level of precision of the
low-dimensional DSR model one needs to, at least, double the
dimensionality of Hilbert space of the system within the simple
truncation approximation.
Finally, assuming the regime of weak HO-bath coupling and low
temperatures, we were able to analytically determine the
dissipative TSS dynamics using a $3$LS Redfield theory.

\section{Acknowledgements}
AOC would like to thank the CNPq (Conselho Nacional de
Desenvolvimento Cient{\'\i}fico e Tecnol{\'o}gico) and the
Millenium Institute for Quantum Information for financial support.

\section*{References}
\bibliographystyle{unsrt}
\bibliography{NJP_revised}
\end{document}